\documentclass[prl,twocolumn,notoofinbib,aps,showpacs,a4paper,floats,
superscriptaddress]{revtex4}

\usepackage{url}
\usepackage{enumerate}

\usepackage[linktocpage,unicode]{hyperref}

\newcommand{\secintro}{I}
\newcommand{\secgauge}{II}
\newcommand{\secbasics}{III}
\newcommand{\secgr}{IV}
\newcommand{\secconclusion}{V}

\def\p{\partial}
\def\Lie{{\cal L}}

\usepackage{mathrsfs}
\usepackage{amsmath,amsfonts,amssymb,amsthm}
\usepackage{comment}

\newtheorem*{Def}{Definition} 
\newtheorem*{lem}{Lemma} 
\newtheorem*{Thm}{Theorem} 

\usepackage{color}

\definecolor{gray}{rgb}{0.8,0.8,0.8} 
\definecolor{cyan}{rgb}{0,0.9,0.9} 
\definecolor{orange}{rgb}{0.9,0.5,0} 
\definecolor{magenta}{rgb}{1,0,1}

\begin{document}


\title{Hyperbolicity of Physical Theories with Application to 
General Relativity}

\author{David Hilditch}
\affiliation{Theoretical Physics Institute, University of 
Jena, 07743 Jena, Germany}
\author{Ronny Richter}
\affiliation{Mathematisches Institut, Universi\"at T\"ubingen, 
72076 T\"ubingen, Germany}

\begin{abstract}
We consider gauge theories from the free evolution point of view, in 
which initial data satisfying constraints of a theory are given. 
Because the constraints are compatible with the field equations 
they remain so. We study a model constrained Hamiltonian theory and 
identify a particular structure in the equations of motion which 
we call the standard gauge freedom. The pure gauge subsystem 
of this model theory is identified and the manner in which the 
gauge variables couple to the field equations is presented. We 
demonstrate that the set of gauge choices that can be coupled to 
the field equations to obtain a, properly defined, wave-like 
formulation is exactly the set of wave-like pure gauges. 
Consequently we analyze a parametrized family of formulations 
of general relativity. The generalization of the harmonic gauge 
formulation to a five parameter family of gauge conditions is 
obtained.
\end{abstract}

\pacs{
  95.30.Sf,   
  04.25.D-   
}

\maketitle

\noindent{\bf{\em \secintro.} Introduction.}
Field theories often have wave-like, or hyperbolic, 
degrees of freedom contained somehow in a set of variables, some 
of which are constrained, and some of which, the gauge fields, 
are not determined by the theory~\cite{Ger96}. Physical states 
are equivalence classes of solutions related by a change of gauge. 
Solutions to the theory can be understood through properties 
of the equations of motion, which consist of a mixture of the 
gauge, constraint and physical quantities. Unraveling this 
structure in general may be hopeless. But if the gauge is 
carefully chosen, say by taking the harmonic gauge in general 
relativity~\cite{Bru52}, then the full set of equations of 
motion may be rendered strongly 
hyperbolic~\cite{GusKreOli95,GunGar05}. This condition 
guarantees the existence of a unique solution to the initial 
value problem that depends continuously on the initial data, at 
least locally in time. As highlighted 
in~\cite{FriRen00}~``{\it Ideally, one would like to 
exhibit a kind of hyperbolic skeleton of the Einstein equations 
and a complete characterization of the freedom to fix the gauge 
from which all hyperbolic reductions should be derivable. Instead, 
there are at present various different methods available which 
have been invented to serve specific needs,}'' this ad-hoc 
characterization is unsatisfactory. Equations of motion for the 
gauge choice can be obtained in the absence of any coupling to 
the theory, which begs the question -- what is 
this skeleton? In other words, what are the set of pure gauges 
that can be coupled to the theory to form a hyperbolic 
formulation? Since the basic characterization of a set of partial 
differential equations can be made in the linear approximation, 
we may start by directing our efforts there. We thus begin to 
address these issues in section~{\secgauge} for a model linear 
constrained Hamiltonian system. In section~{\secbasics} we examine 
conditions under which a formulation of the Hamiltonian theory is 
strongly hyperbolic. In section~{\secgr} we apply our findings to 
general relativity~(GR) with a five parameter family of gauge 
conditions and obtain the generalization of the harmonic 
formulation to this family.

\noindent{\bf{\em \secgauge.} A model theory with gauge 
freedom.}
Consider the equations of motion for the Hamiltonian density,
\begin{align}
\label{eqn:Hamiltonian}
H&=\frac{1}{2}
\left(\begin{array}{c}
\p_iq \\ p
\end{array}\right)^\dagger
\left(\begin{array}{cc}
 V^{ij} & F^{\dagger\,i} \\
 F^{j}  & M^{-1}
\end{array}\right)
\left(\begin{array}{c}
\p_jq \\ p
\end{array}\right)\nonumber\\
&\quad+g_q^\dagger C_{\mathcal H}V^{ij}\p_i\p_jq
+g_p^\dagger C_{\mathcal M}{}^iM^{-1}\p_ip\,,
\end{align}
with canonical positions and momenta~$(q,p)$. Every matrix is 
constant,~$M^{-1}$ is invertible and~$F^i=\beta^i\,I$ for some 
shift vector~$\beta^i$, with~$I$ the appropriate identity. 
Such a Hamiltonian can be obtained from that of GR by 
linearizing~\cite{Mon74} and discarding lower derivatives. 
Variation with respect to the gauge fields~$(g_q,g_p)$ reveals 
the constraints
\begin{align}
{\mathcal H}&=C_{\mathcal H}V^{ij}\p_i\p_jq=0\,,&\quad
{\mathcal M}&=C_{\mathcal M}{}^iM^{-1}\p_ip=0\,,\nonumber
\end{align}
which we will take to be first class and call the Hamiltonian 
and momentum constraints respectively.

\paragraph*{Gauge invariance:} The constraints generate the 
gauge transformation,
\begin{align}
q&\to\bar{q}=q-M^{-1}C_{\mathcal M}{}^{\dagger\,i}\p_i\psi\,,
\nonumber\\
p&\to\bar{p}=p-V^{ij}C_{\mathcal H}{}^\dagger\p_i\p_j\theta\,,
\label{eqn:gauge_transform}
\end{align}
with unspecified fields~$\theta$ and~$\psi$. We require that 
the {\it field strength}~$V^{ij}\p_i\p_jq$ and {\it curl}~$\epsilon^i
M^{-1}\p_ip$, defined by some square
anti-hermitian matrices~$\epsilon^i$, are invariant under this 
transformation. Gauge invariance thus gives,
\begin{align}
&(A_{\mathcal{HM}})^{(i}C_{\mathcal{M}}{}^{j)}=C_{\mathcal H}V^{ij}\,,&\quad
&V^{(ij}M^{-1}C_{\mathcal{M}}{}^{\dagger\,k)}=0\,,\nonumber\\
&\epsilon^{(i}M^{-1}V^{jk)}C_{\mathcal H}{}^\dagger=0\,,\nonumber
\end{align}
for some matrices~$(A_{\mathcal{HM}})^{i}$, where the index 
parentheses denote symmetrization. Gauge invariance of the evolution
equations also implies that
\begin{align}
\p_t\theta&=\beta^i\p_i\theta+\bar{g}_q\,,
\nonumber\\
\p_t\psi&=(A_{\mathcal{HM}})^{\dagger\,i}\p_i\theta
+\beta^i\p_i\psi+\bar{g}_p\,,\label{eqn:pure_gauge}
\end{align}
where~$\bar{g}_q,\bar{g}_p$ denote the change 
under~\eqref{eqn:gauge_transform}, and where here, and in what 
follows,~$\beta^i$ is taken to include the appropriate identity.

\paragraph*{Electric and magnetic degrees of freedom:} Without 
loss of generality the matrix~$C_{\mathcal H}$ has linearly 
independent rows, so we can decompose the potential 
matrix~$V^{ij}$ according to 
\begin{align}
\label{eqn:phys_proj}
V^{ij}&=V_{\mathcal P}^{ij}
+\hat{C}_{\mathcal H}{}^\dagger\hat{C}_{\mathcal H}V^{ij},&\quad
V_{\mathcal P}^{ij}=\,\perp_{\mathcal{P}} V^{ij}\,,
\end{align}
with~$\hat{C}_{\mathcal H}=[C_{\mathcal H}C_{\mathcal H}{}^\dagger
]^{-1/2}C_{\mathcal H}$, and the projection operator~$\perp_{\mathcal{P}}$ 
defined implicitly by~\eqref{eqn:phys_proj}.
In the absence of Hamiltonian 
constraints we assume that~$V^{ij}=V_{\mathcal P}^{ij}$, and 
always that~$V_{\mathcal P}^{ij}=\epsilon^{\dagger\,i}\,(A_{VB})\,
\epsilon^j+C_{\mathcal{M}}{}^{\dagger\,(i}(A_{V\mathcal{M}})C_{\mathcal{M}}{}^{j)}$ for 
some Hermitian matrices~$(A_{VB})$ and~$(A_{V\mathcal{M}})$. The electric 
and magnetic degrees of freedom are,
\begin{align}
E&=V_{\mathcal P}^{ij}\p_i\p_jq\,,&\quad 
B=\epsilon^iM^{-1}\p_ip\,.\nonumber
\end{align}
which, up to coupling to the constraints, form a closed subsystem 
and are gauge invariant. Such fields can be similarly defined 
in the absence of Hamiltonian constraints. They are not used 
in the analysis that follows.

\paragraph*{Closure of the pure gauge subsystem:} We call an 
equation of motion for the gauge fields a gauge choice. Here 
we consider only evolution conditions
\begin{align}
\p_tg_q&=(A_{g_qg_q})^i\p_ig_q+(A_{g_qg_p})^i\p_ig_p
+(A_{g_qp})p\,,\nonumber\\
\p_tg_p&=(A_{g_pg_q})^i\p_ig_q+(A_{g_pg_p})^i\p_ig_p
+(A_{g_pq})^i\p_iq\,.
\label{eqn:gauge_choice}
\end{align}
We assume that~$(A_{g_qp})=AC_{\mathcal H}$ 
and~$(A_{g_pq})^i=BC_{\mathcal{M}}{}^{i}+C^iC_{\mathcal H}M$ for some 
matrices~$A,B$ and~$C^i$, a restriction which can be dropped 
by altering our arguments slightly. Assume that we are given a 
solution to the theory. We have already seen that the field equations 
are invariant under the gauge 
transformation~\eqref{eqn:gauge_transform}. The pure gauge 
subsystem~\eqref{eqn:pure_gauge} is closed by substituting the 
gauge difference from~\eqref{eqn:gauge_transform} 
into~\eqref{eqn:gauge_choice}, taking~$g_q\to\bar{g}_q$ 
and~$g_p\to\bar{g}_p$.

\paragraph*{Free evolution on the expanded phase space:} We 
are free to modify the dynamics of the model theory away from
the constraint satisfying hypersurface in phase space, provided 
that the constraint subsystem remains closed. We  define new 
constraints~$(\Theta,Z)$ with the same 
length as~$(g_q,g_p)$ respectively. We couple the new 
constraints to the gauge conditions~\eqref{eqn:pure_gauge} 
by parametrized addition according to
\begin{align}
\p_tg_q&=(A_{g_qg_q})^i\p_ig_q+(A_{g_qg_p})^i\p_ig_p
+(A_{g_qp})p\nonumber\\
&\quad+(A_{g_q\Theta})\Theta\,,\nonumber\\
\p_tg_p&=(A_{g_pg_q})^i\p_ig_q+(A_{g_pg_p})^i\p_ig_p
+(A_{g_pq})^i\p_iq\nonumber\\
&\quad+(A_{g_pZ})Z\,.\nonumber
\end{align}
Likewise for the equations of motion
\begin{align}
\p_tq&=M^{-1}p+F^i\p_iq-M^{-1}C_{\mathcal M}{}^{\dagger\,i}\p_ig_p
+(A_{q\Theta})\Theta\,,\nonumber\\
\p_tp&=V^{ij}\p_i\p_jq+F^i\p_ip-V^{ij}C_{\mathcal H}{}^\dagger\p_i\p_jg_q
+(A_{pZ})^i\p_iZ\nonumber\\
&\quad+(A_{p\mathcal{H}})\mathcal{H}\,.\nonumber
\end{align}
We choose equations of motion for the new constraints
\begin{align}
\p_t\Theta&=\beta^i\p_i\Theta
+(A_{\Theta Z})^i\p_iZ
+(A_{\Theta\mathcal{H}})\mathcal{H}\,,\nonumber\\
\p_tZ&=(A_{Z\Theta})^i\p_i\Theta
+\beta^i\p_iZ+(A_{Z\mathcal{M}})\mathcal{M}\,.\nonumber
\end{align}
The constraint subsystem is closed by 
\begin{align}
\p_t\mathcal{H}&=(A_{\mathcal{H}\Theta})^{ij}
\p_i\p_j\Theta+\beta^i\p_i\mathcal{H}
+(A_{\mathcal{HM}})^i\p_i\mathcal{M}\,,\nonumber\\
\p_t\mathcal{M}&=(A_{\mathcal{M}Z})^{ij}\p_i\p_jZ
+(A_{\mathcal{MH}})^i\p_i\mathcal{H}+\beta^i\p_i\mathcal{M}\,,\nonumber
\end{align}
with matrices
\begin{align}
(A_{\mathcal{H}\Theta})^{ij}&=
C_{\mathcal H}V^{ij}(A_{q\Theta})\,,\nonumber\\
(A_{\mathcal{M}Z})^{ij}&=C_{\mathcal M}{}^{(i}M^{-1}
(A_{pZ})^{j)}\,,\nonumber\\
(A_{\mathcal{M}{\mathcal{H}}})^i&=C_{\mathcal M}{}^iM^{-1}
(A_{p\mathcal{H}})\,.\nonumber
\end{align}

\paragraph*{Natural choice of variables:} The next assumption
is that the variables can be appropriately broken up. For this 
we assume that for every unit spatial vector~$s^i$, the rows 
of~$C_{\mathcal{H}}{}$ and~$C_{\mathcal{M}}{}^{s}\equiv 
C_{\mathcal{M}}{}^{i}s_i$ are contained in the span of the union of 
the rows of~$V=C_{\mathcal{H}}V^{ss}$ and~$W=C_{\mathcal{M}}{}^sM^{-1}$, 
which each have themselves independent rows, and furthermore 
that the contractions~$X=VC_{\mathcal{H}}{}^\dagger$ 
and~$Y=WC_{\mathcal{M}}{}^{\dagger s}$ are invertible. With these 
conditions we can define 
\begin{align}
C_{\theta}&=-X^{-1}C_{\mathcal{H}}\,,\nonumber\\
C_{\psi}&=-Y^{-1}C_{\mathcal{M}}{}^s
+(A_{\mathcal{HM}})^{\dagger s}C_{\theta}[M
-C_{\mathcal{M}}{}^{\dagger s}Y^{-1}
C_{\mathcal{M}}{}^s]\,,\nonumber\\
\perp&=I-V^\dagger[VV^\dagger]^{-1}V-W^\dagger[WW^\dagger]^{-1}W\,,
\nonumber
\end{align}
and the decomposition of~$\p_sq$ and~$p$ into gauge, constraint, 
and physical degrees of freedom,
\begin{align}
\p_s^2\theta&=C_{\theta}p
+(A_{\theta\Theta})\Theta\,,&\quad
\p_s^2\psi&=C_{\psi}\p_sq+(A_{\psi Z})Z\,,\nonumber\\
\mathcal{H}&=V\p_sq\,,&\quad
\mathcal{M}&=Wp\,,\nonumber\\
\p_sP_q&=\,\perp\p_sq\,,&\quad
P_p&=\,\perp p\,,\nonumber
\end{align}
is invertible. The names here serve only to 
identify the relationship between the pure gauge and 
constraints.

\paragraph*{Principal symbol of a formulation:} Once the gauge 
and constraint addition parameters are fixed we say that we 
have a formulation of the theory. The principal symbol of a 
formulation in the~$s^i$ direction is
\begin{align}
\label{eqn:Symbol}
P^s&=\left(
\begin{array}{ccc} 
P_{\mathcal G}^s & P_{\mathcal{GC}}^s & 0 \\
 0 & P_{\mathcal C}^s & 0 \\
 0 & 0 & P_{\mathcal P}^s \\
\end{array}\right)\,.
\end{align}
We assume that the constraint addition parameters are annihilated 
by the projection operator~$\perp$. This restriction can also 
be relaxed. The pure gauge sub-block, 
\begin{align}
P_{\mathcal{G}}^s&=
\left(\begin{array}{cccc}
\beta^s & 0 & I & 0 \\
(A_{\mathcal{HM}})^{\dagger s} & \beta^s & 0 & I \\
-(A_{g_qp})V^\dagger & 0 & (A_{g_qg_q})^s & (A_{g_qg_p})^{s}\\
0 & -(A_{g_pq})^sW^\dagger & (A_{g_pg_q})^s & (A_{g_pg_p})^s
\end{array}\right),\nonumber
\end{align}
is exactly the principal symbol of the pure gauge subsystem described after 
equation~\eqref{eqn:gauge_choice}. 
The off-diagonal block,
\begin{align}
P_{\mathcal{GC}}^s&=
\left(\begin{array}{cccc}
0 & (A_{\theta Z}) & (A_{\theta\mathcal{H}}) & 0 \\
(A_{\psi\Theta}) & 0 & 0 & (A_{\psi\mathcal{M}}) \\
(A_{\Theta}) & 0 & 0 & 0 \\
0 & (A_{Z}) & 0 & 0
\end{array}\right),\nonumber
\end{align}
with sub-matrices,
\begin{align}
(A_{\theta Z})&=(A_{\theta\Theta})(A_{\Theta Z})+C_{\theta}(A_{pZ})^s\,,\nonumber\\
(A_{\theta\mathcal{H}})&=(A_{\theta\Theta})(A_{\Theta\mathcal{H}})-X^{-1}+C_{\theta}(A_{p\mathcal H})
\,,\nonumber\\
(A_{\psi\Theta})&=(A_{\psi Z})(A_{\psi\Theta})-(A_{\mathcal{HM}})^{\dagger s}(A_{\theta\Theta})
+C_{\psi}(A_{g_q\Theta})
\,,\nonumber\\
(A_{\psi\mathcal{M}})&=(A_{\psi Z})(A_{Z\mathcal{M}})-Y^{-1}
-(A_{\mathcal{HM}})^{\dagger s}C_{\theta}
C_{\mathcal{M}}{}^{\dagger s}Y^{-1}\,,\nonumber\\
(A_{\Theta})&=(A_{g_qp})V^\dagger(A_{\theta\Theta})+(A_{g_q\Theta})\,,\nonumber\\
(A_{Z})&=(A_{g_pq})^sW^\dagger (A_{\psi Z})+(A_{g_pZ})
\,,\nonumber
\end{align}
parametrizes the coupling of the gauge fields to the constraints. 
The constraint violating sub-block,
\begin{align}
P_{\mathcal{C}}^s&=
\left(\begin{array}{cccc}
\beta^s & (A_{\Theta Z})^s & 
(A_{\Theta\mathcal{H}}) & 0\\
(A_{Z\Theta})^s & \beta^s & 0 & 
(A_{Z\mathcal{M}})\\
(A_{\mathcal{H}\Theta})^{ss} & 0 & \beta^s & 
(A_{\mathcal{HM}})^s\\
0 & (A_{\mathcal{M}Z})^{ss} & 
(A_{\mathcal{MH}})^s & \beta^s
\end{array}\right),\nonumber
\end{align}
is exactly the principal symbol of the constraint subsystem.
Finally the physical sub-block,
\begin{align}
P_{\mathcal P}^s&=
\left(\begin{array}{cc}
\beta^s&\perp M^{-1}\\
\perp V^{ss}&\beta^s
\end{array}\right)\,,\nonumber
\end{align}
contains neither constraint addition or gauge parameters.

\paragraph*{Strong hyperbolicity:} A necessary condition for 
strongly hyperbolicity is that~$P^s$ has real eigenvalues and 
a complete set of eigenvectors for every~$s^i$. Strong 
hyperbolicity is equivalent to well-posedness, that is existence 
of a unique solution depending continuously on the given data, 
of the initial value problem~\cite{GusKreOli95,GunGar05,SarTig12}.

\noindent{\bf{\em \secbasics.} Basic properties of theories 
with the standard gauge freedom.}
Consider the theory of the previous section. Then:

\begin{lem} No formulation is strongly hyperbolic if the 
physical sub-block is not.
\begin{proof}
Obviously a necessary condition for diagonalizability with 
real eigenvalues of~\eqref{eqn:Symbol}, for any formulation, 
is that of~$P_{\mathcal P}^s$.
\end{proof}
\end{lem}

\begin{lem} 
A necessary condition for strong hyperbolicity of a formulation
is that the pure gauge and constraint violating subsystems are 
strongly hyperbolic.
\begin{proof}
We need to show that if the matrix~\eqref{eqn:Symbol} is 
diagonalizable with real eigenvalues then this property holds 
for the pure gauge and constraint violating sub-blocks. A 
diagonalizable upper block triangular matrix has diagonalizable 
blocks on the diagonal~\cite[App. A]{GunGar05}. Moreover, the 
set of eigenvalues of the full matrix is the union of the 
eigenvalues of the diagonal blocks. The lemma follows.
\end{proof}
\end{lem}

\begin{Def} If for every strongly hyperbolic pure gauge 
there exists a choice of constraint addition parameters so that 
the formulation is strongly hyperbolic, we say that the theory has 
the standard gauge freedom. 
\end{Def}

\begin{Thm} Given a theory with the standard gauge freedom, a pure 
gauge can be used to form a strongly hyperbolic formulation if and 
only if it is strongly hyperbolic.
\begin{proof}
The result follows trivially from the Lemmas and the 
definition of the standard gauge freedom.
\end{proof}
\end{Thm}

\noindent{\bf{\em \secgr.} Application to GR.}
The ADM Hamiltonian~\cite{ArnDesMis62} for vacuum 
GR is~$H_{\textrm{ADM}}=-\alpha\,H+2\,\beta^i\,M_i,$
with Hamiltonian and momentum constraints
\begin{align}
H&=R-K_{ij}K^{ij}+K^2\,,&\quad M_i=D^jK_{ij}-D_iK\,.\nonumber
\end{align}

\paragraph*{Gauge freedom in the non-linear regime:} We 
take the freedom to be to choose coordinates~$x^\mu=(t,x^i)$ 
on spacetime; qualitative features of the model carry over. 
The constraints are obviously spatially covariant. Given 
an additional {\it upper case} time coordinate~$T$ with normal 
vector~$N^a$ such that~$N^a=W(n^a+v^a),$ with Lorentz factor~$W$ 
and spatial boost vector~$v^i$ then
\begin{align}
{}^{(N)}H&=W^2H-2W^2M_v\,,\nonumber\\
\perp\cdot{}^{(N)}M_i&=WM_i+2W^3M_vv_i-W^3Hv_i\,,\nonumber
\end{align}
where~$\perp^a_b$ is the projection operator into slices of 
constant~$t$, and subscript~$v$ denotes contraction with the 
velocity~$v^i$. The electric and magnetic parts of the Weyl 
tensor~\cite{MaaBas97}, form a closed subsystem, up to coupling 
to the constraints, and from the point of view of the lower case 
observer the spatial part of the upper case electric and magnetic 
parts are
\begin{align}
\perp\cdot{}^{(N)}E_{ij}&=(2W^2-1)E_{ij}-2W^2E_{v(i}v_{j)}
+W^2E_{vv}\gamma_{ij}\nonumber\\
&\qquad+2W^2\epsilon^k{}_{v(i}B_{j)k}
\,,\nonumber\\
\perp\cdot{}^{(N)}B_{ij}&=W^2B_{ij}-W^2\epsilon^k{}_{ij}E_{kv}
-W^2\epsilon^k{}_{vi}E_{jk}\,,\nonumber
\end{align}
which shows that if the fields vanish in one foliation they 
vanish in every foliation.

\paragraph*{Linearized pure gauge subsystem:} The linearized 
pure gauge subsystem is~\cite{KhoNov02},
\begin{align}
\p_t\theta&=U-\psi_iD^i\alpha+\beta^i\p_i\theta,\nonumber\\
\p_t\psi^i&=V^i+\alpha D^i\theta-\theta D^i\alpha+\Lie_\beta\psi^i,
\label{eqn:GR_pure}
\end{align}
where~$\theta=-n_a\Delta[x^a]$,~$\psi^i=-\perp^i_a\Delta[x^a]$, 
$U=\Delta[\alpha]$ and $V^i=\Delta[\beta^i]$. Under 
an infinitesimal change of gauge the perturbation to the metric 
and extrinsic curvature are given by the York 
equations~\cite{Yor79a} with~$\alpha\to\theta$ 
and~$\beta^i\to\psi^i$, which can be used to close the linearized 
pure gauge subsystem once we act on the gauge condition with 
the perturbation operator~$\Delta$.

\paragraph*{Free evolution in the expanded phase space:} 
We expand the phase space by constraints~$\Theta$ 
and~$Z_i$, and parametrize the equations of motion for the 
gauge by
\begin{align}
\p_t\alpha &= - g_1\alpha^2K + g_2\alpha\p_i\beta^i
+\beta^i\p_i\alpha+2c_1\alpha^2\Theta\,,\nonumber\\
\p_t\beta^i &= \alpha^2[g_3\gamma^{kl}\gamma^{ij}
               + g_4\gamma^{il}\gamma^{jk}]\p_l\gamma_{jk}
               - g_5\alpha\p^i\alpha
               + \beta^j\p_j\beta^i\nonumber\\
&\qquad+2\alpha^2c_2Z^i\,,\label{eqn:GR_Gauge}
\end{align}
with~$g_1>0$ and~$\bar{g}_3=2(g_3+g_4)>0$, and for 
the remaining variables by
\begin{align}
\p_t\gamma_{ij}&=-2\alpha K_{ij}+\Lie_\beta\gamma_{ij}
+\tfrac{1}{3}\,c_3
\alpha\gamma_{ij}\Theta\,,\nonumber\\
\p_t K_{ij}&= -D_iD_j\alpha+\alpha[R_{ij}-2K^k{}_iK_{jk}+K_{ij}K]
+\Lie_\beta K_{ij}\nonumber\\
&\qquad+2c_4\alpha \p_{(i}Z_{j)}
+\tfrac{1}{3}\,c_5 \alpha\gamma_{ij}\p_kZ^k
+\tfrac{1}{3}\,c_6\alpha\gamma_{ij}H\,,\nonumber\\
\p_t\Theta&= c_7\alpha H+c_8\alpha\p_iZ^i+\Lie_\beta\Theta\,,
\nonumber\\
\p_tZ_i&= c_9\alpha M_i+c_{10} \alpha \p_i\Theta + \Lie_\beta Z_i\,. 
\nonumber
\end{align}
Strong hyperbolicity for non-linear and variable coefficient 
systems is defined, with additional smoothness conditions, by 
linearizing and working in the high-frequency frozen coefficient 
approximation~\cite{SarTig12}. In this approximation the 
Hamiltonian density~\cite{Mon74} has the structure 
of~\eqref{eqn:Hamiltonian}. 

\paragraph*{Strong hyperbolicity of the pure gauge subsystem:}
The principal symbol of the linearized pure gauge 
subsystem~\eqref{eqn:GR_pure} with gauge 
choice~\eqref{eqn:GR_Gauge}, where one must ignore constraint 
addition, has eigenvalues~$\pm\sqrt{g_3}\,,\pm v_{\pm},$ with
\begin{align}
2\,v_{\pm}^2&=g_1+\bar{g}_3-g_2g_5\nonumber\\
&\quad\pm\sqrt{(g_1+\bar{g}_3-g_2g_5)^2-4(g_1-g_2)\bar{g}_3}\,.
\nonumber
\end{align}
The subsystem is strongly hyperbolic if~$g_3>0$ and either 
\begin{enumerate}[i).]
\item $0\ne g_2<g_1$ 
and~$g_2g_5<g_1-2\sqrt{g_1-g_2}\sqrt{\bar{g}_3}+\bar{g}_3$\,,
\item $g_2=0$ and~$\bar{g}_3\ne g_1$ or $g_2=0,\bar{g}_3=g_1$
and $g_5=1$\,.
\end{enumerate}
the second clause of case ii). is that of generically 
distinct eigenvalues colliding without loss of diagonalizability.  

\paragraph*{Strong hyperbolicity of the constraint subsystem 
with vanishing gauge-constraint coupling:} Choosing
\begin{align}
c_1&=g_1\,,\quad c_2=g_3\,,\quad c_3=c_5=c_6=0\,, 
\nonumber\\
c_4&=2\,c_7=c_8=c_9=1\,,\quad
c_{10}=2\Big(1+\frac{g_4}{g_3}\Big)\,,\label{eqn:formulation}
\end{align}
guarantees both that the off-diagonal block of the principal 
symbol~$P_{\mathcal{GC}}^s$ vanishes and that the constraint 
subsystem is strongly hyperbolic. The eigenvalues of the 
constraint violating sub-block~$P_{\mathcal{C}}^s$ 
are~$\pm\sqrt{c_{10}}$, which are guaranteed to be real inside 
the class of gauges we are considering, and~$\pm 1$ with 
multiplicity three. 

\paragraph*{Strong hyperbolicity of physical sub-block:} The 
physical sub-block is diagonalizable with eigenvalues~$\pm1$, 
at least up to a trivial normalization. Assuming 
smoothness of the background implies the continuity requirement 
for strong hyperbolicity in every block.

\paragraph*{Discussion:} The choice~\eqref{eqn:formulation} is 
the natural extension of the harmonic gauge 
formulation~\cite{Bru52} to the family of gauge 
conditions~\eqref{eqn:GR_Gauge}. If a gauge in which the 
contracted Christoffel symbol is chosen to appear in the shift 
condition, i.e when~$g_4=-\tfrac{1}{2}g_3$, the constraint 
addition parameters correspond to those of the principal part 
of the Z4 formulation~\cite{BonLedPal03}. Otherwise it differs 
in the constraint subsystem.

\noindent{\bf{\em \secconclusion.} Conclusion.}
Stimulated by~\cite{FriRen00}, in which the possibility 
of identifying {\it every} hyperbolic formulation 
of GR was suggested, we identified a particular structure 
in constrained Hamiltonian equations of motion. We examined how 
pure gauge is inherited by a formulation of a theory. With this 
structure the set of strongly hyperbolic pure gauges are exactly 
those that can be used to form a strongly hyperbolic formulation, 
in-line with the expectation of the physicist. We expect that 
the results can be generalized to include elliptic gauges. It will 
furthermore be of interest to treat the initial boundary value 
problem. We used our findings to investigate hyperbolicity of 
a family of formulations of GR, generalizing~\cite{HilRic10} to 
non Hamiltonian formulations. Open questions include those 
relating to long-term existence with different gauges. 

\noindent{\bf{Acknowledgments.}}
The authors thank S. Bernuzzi, B. Br\"ugmann, C. Gundlach, 
N. \'O~Murchada, M. Ruiz and A. Weyhausen for helpful 
discussions. The work was partially supported by DFG grant 
SFB/Transregio~7. 

\bibliographystyle{apsrev}
\bibliography{Gauge.bbl}{}


\end{document}